\documentclass{appolb}
\usepackage{graphicx}
\usepackage{amsmath}
\usepackage{amssymb}
\usepackage{lineno}
\usepackage{hyperref}
\def\Jpsi{\ensuremath{\text{J}/\psi}\xspace}

\begin{document}
\title{Recent ALICE results relevant for PDFs
at low and high-$x$, saturation%
\thanks{Presented at ``Diffraction and Low-$x$ 2024'', Trabia (Palermo, Italy), September 8-14, 2024.}%
}
\author{Anisa Khatun for the ALICE Collaboration
\address{The University of Kansas}
\\[3mm]
}

\maketitle
\begin{abstract}
We present recent results from the ALICE Collaboration on the study of coherent and incoherent J/$\psi$ photoproduction in ultra-peripheral Pb-Pb collisions, including results from exclusive and dissociate J/$\psi$ mesons in ultra-peripheral p-Pb interactions. These measurements provide unique insights into the initial state of protons and ions, with great sensitivity for both gluon saturation and shadowing. Furthermore, we will discuss the prospects for these measurements using the Run 3 and Run 4 data.
\end{abstract}

\section{Introduction}

Heavy-ion collisions recreate a hot and dense, deconfined state of nuclear matter known as the Quark-Gluon Plasma (QGP). In contrast, small systems like proton-proton, proton-nucleus, and $\gamma$-nucleus collisions are not expected to form a hot, thermalized medium. Nevertheless, some QGP-like signatures, such as the suppression of quarkonium, can occur due to medium-induced modifications. These phenomena are referred to as cold nuclear matter (CNM) effects.
Disentangling CNM effects from the QGP is crucial for accurately interpreting heavy-ion collision data. Additionally, CNM effects play a significant role in understanding nucleon and nucleus structure. 

The CNM effects are quantified using the nuclear modification factor: $R_{iA}(x,Q^{2})
= (1/A)f_{iA}(x,Q^{2})/f_{i}(x,Q^{2})$, where $f_{iA}(x,Q^{2})$ is the Parton Distribution Function (PDF) of a parton flavour “$i$” inside a proton bound in a nucleus, and $f_{i}(x,Q^{2})$ is the PDF of a free proton. A is the number of nucleons of the target nucleus. The PDF depends on both Bjorken$-x$ and momentum transfer scale $Q^{2}$. A value of $R_{iA}(x,Q^{2})= 1$ indicates the absence of nuclear effect~\cite{EPhJC822022413}.

At low Bjorken$-x$ the gluon density increases with decreasing $x$ and eventually saturates~\cite{EPCJ752015580}. Two CNM effects of particular interest at LHC energies are gluon saturation and gluon shadowing. These phenomena can be explored through vector meson photoproduction in ultra-peripheral collisions (UPCs), which provide access to extremely low Bjorken$-x$ values (down to approximately $10^{-6}$ at LHC energies, Eqn.~\ref{eqn1}). UPCs thus serve as a valuable tool to probe the structure of nuclei and nucleons.

\section{Physics of UPCs}
\label{section2}

In ultra-peripheral collisions (UPCs), ions interact via electromagnetic processes, with no hadronic interactions taking place. This occurs when two relativistic heavy ions interact with an impact parameter larger than the sum of the radii of the colliding nuclei. Experimentally, UPC events are characterized by only a few tracks in an otherwise empty detector, a feature that applies specifically to exclusive (double-gap) events. Examples include exclusive (Pb-Pb) and dissociative (p-Pb) vector meson photoproduction. In the exclusive process, both the projectile and the target nuclei remain intact. This process can be further categorized into two sub-classes: coherent process (photon interacts with the entire nucleus) and incoherent (where the photon interacts with individual nucleons or groups of nucleons within the nucleus).

This article discusses photoproduction cross-section measurements as a function of rapidity ($y$), transverse momentum (Mandelstam$-|t|$) and photon-nucleon center-of-mass energy ($W_{ \gamma \rm A}$). These variables are interrelated and also connect to Bjorken$-x$ as shown in equations \ref{eqn1} - \ref{eqn3}.

\begin{equation}
x = \frac{M_{\rm VM}}{\sqrt{{s}_{NN}}} e^{\pm y} = \frac  {M^{2}_{\rm VM}} {W^{2}_{ \gamma \rm A} }
\label{eqn1}
\end{equation}


\begin{equation}
\text{Mandelstam} - |t| \approx - p^{2}_{\text T},
\label{eqn3}
\end{equation}

\noindent where $M_{\rm VM}$ is the invariant mass of the vector meson and $\sqrt{s_{NN}}$ describes the center-of-mass energy of the Pb-Pb collisions system per nucleon pair. 

\section{UPC Studies in ALICE}
The ALICE experiment has delivered extensive physics results over the past decade, with unique capabilities like wide acceptance for low-momentum particles and excellent midrapidity particle identification~\cite{Alice2}. ALICE facilitates UPC studies at both mid and forward rapidity, leveraging its advanced detectors during Run 2. 

At midrapidity ($|\eta| < 0.9, |y| < 0.8 \rightarrow x \sim 10^{-3}$) key detectors include the Inner Tracking System (ITS) for tracking and vertexing, the Time Projection Chamber (TPC) for tracking and particle identification (PID), and the Time-Of-Flight (TOF) for triggering and PID. The V0 and T0 detectors are used for triggering and vetoing, while the Zero Degree Calorimeters (ZDC) play a crucial role in vetoing events involving neutron and proton emissions.

At forward rapidity ($-4.0 < \eta < - 2.5$ and $-4.0 < y < - 2.5 \rightarrow x \sim 10^{-2} \ \rm{or} \ 10^{-5}$), the Muon Tracker (MCH) and Muon Trigger (MTR) are used for muon tracking and triggering, respectively. The V0 and ALICE Diffractive (AD) detectors are employed for vetoing diffractive events. The ongoing Run 3 program introduces continuous readout and upgraded detectors, enabling new physics studies~\cite{Aliceupgrade, Alic3upc}.

\section{Coherent and exclusive \Jpsi measurements with ALICE}
The exclusive UPC processes in p-Pb and Pb-Pb collisions can be classified into sub processes as explained in Section~\ref{section2}. 
Experimentally, coherent and incoherent processes are disentangled by a cut on transverse momentum $p_{\rm T} \lesssim $ 0.2 GeV/c (coherent) or $p_{\rm T} \gtrsim $ 0.2 GeV/c (incoherent).

Differential photonuclear cross-section of coherent J/$\psi$ is measured in terms of $\it{y}$ via its dimuon decay channel. The results are compared with various model predictions like Impulse approximation (no nuclear effects)~\cite{IAmodel}, STARlight (hadronic model based on Glauber calculations)~\cite{Starlight}, EPOS09 LO (parametrization of available nuclear shadowing data)~\cite{EPOS09}, LTA (leading twist approximation) of nuclear shadowing using Gribov-Glauber theory~\cite{LTA}, GG-HS – color dipole model + gluon saturation (hot spot model for hadronic structure)~\cite{GGHS}, The b-BK-A model~\cite{bbka}, based on a solution of the impact-parameter dependent Balitsky-Kovchegov (BK) equation (valid only at low-$x<10^{-2}$) etc, shows an indication of gluon shadowing~\cite{PLB798}. 

\begin{figure}[htb]
\includegraphics[width=0.50\textwidth]{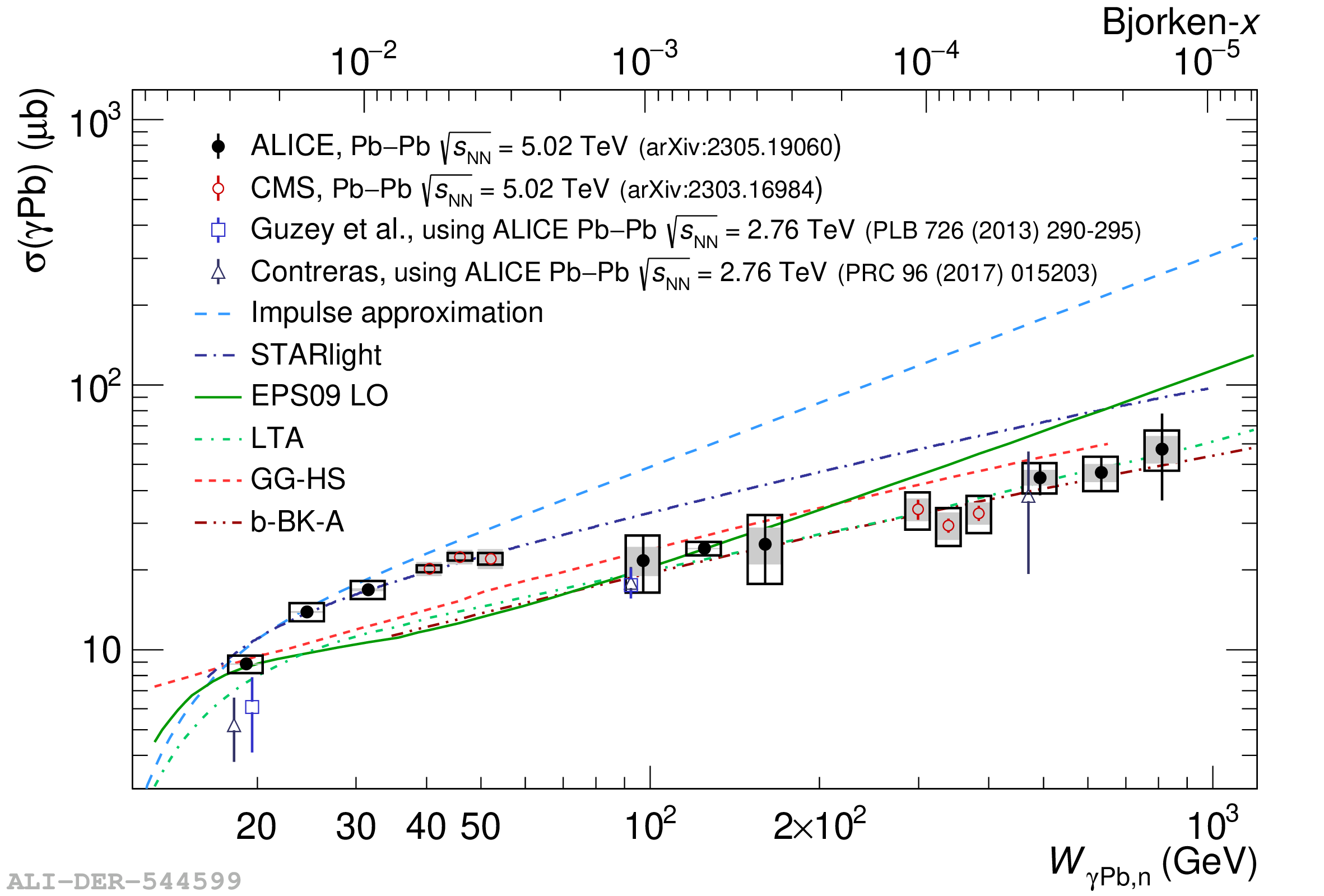}
\includegraphics[width=0.51\textwidth]{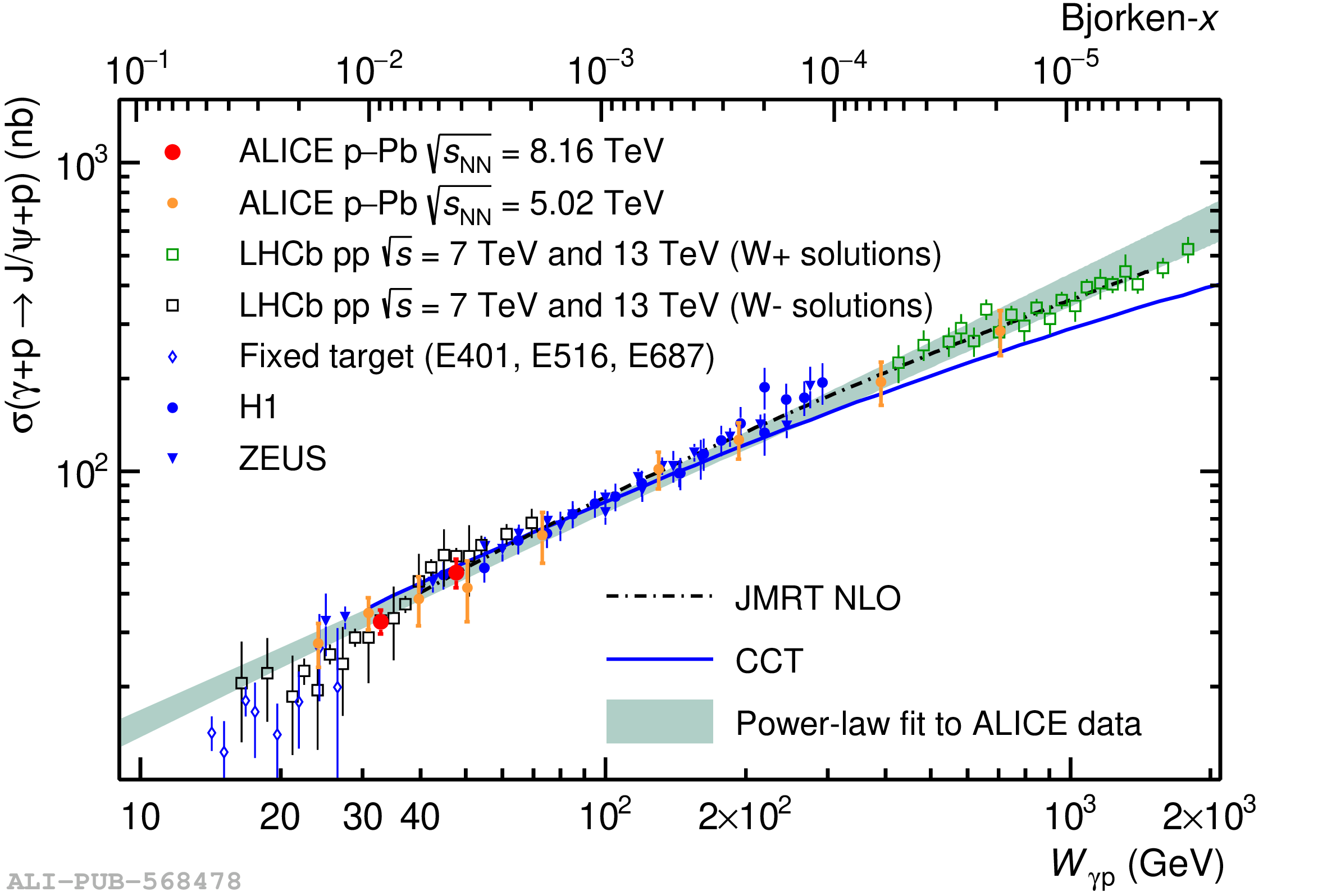}
\caption{(a). Energy dependence of coherent (lead target)~\cite{JHEP10} and (b). exclusive (proton target)  J/$\psi$ cross-section compared with various theoretical models~\cite{H1, HERA}.}
\label{Fig:energydep}
\end{figure}



The UPC cross section is expressed as the product of the photon flux and the photonuclear cross section($\frac{\rm{d}\sigma_{\rm PbPb J/\psi}}{{\rm d} y}$), as shown in Eq.~\ref{eq:upc_cs}. At midrapidity, the contributions from the photon source and target are equal, while at forward rapidity, they differ. These contributions can be disentangled using neutron emission classes, which correspond to different impact parameters (b) and electromagnetic dissociation (EMD) of nuclei via independent photon exchange. XnXn Neutrons are emitted on both sides of the beam, corresponding to small b range. 0nXn or Xn0n: Neutrons are emitted on only one side, indicating medium b range. 0n0n: No neutrons are detected, associated with large b~\cite{PRL89, EPJC74}.

\begin{equation}
    \frac{\rm{d}\sigma_{\rm PbPb J/\psi}}{{\rm d} y} = n_{\gamma}(y) \sigma_{\gamma{\rm Pb}}(y) + n_{\gamma}(-y) \sigma_{\gamma{\rm Pb}}(-y) \, 
\label{eq:upc_cs}
\end{equation}

Energy dependence of coherent photonuclear cross section and nuclear suppression factor are measured in UPC Pb-Pb collisions at $\sqrt{s_{NN}}$ = 5.02 TeV using the above mentioned approach~\cite{JHEP10}. As shown in Fig~\ref{Fig:energydep} (a), results includes both mid and forward rapidity data. The data covers $17 < W_{\gamma \rm Pb} < 920$ GeV with Bjorken$-x$ interval $1.1\times 10^{-5} < x < 3.3\times 10^{-2}$. No model describe the data completely. STARLight~\cite{Starlight} and the impulse approximation~\cite{IAmodel} describe well low energy data, at high energy data are described by both models that include shadowing (EPS09-LO~\cite{EPOS09}, LTA~\cite{LTA}) and saturation effects (GG-HS)\cite{GGHS}. 

In p-Pb collisions, the system is asymmetric, eliminating the source/target ambiguity in $W_{ \gamma \rm p}$ energy since the dominant source of photons is the Pb-nucleus. ALICE covers a broad energy range at 5.02 and 8.16 TeV, with $ 20 < W_{\gamma \rm p}  < 70$ GeV. The low-energy photons are emitted predominantly by the nucleus (Pb). 
Exclusive J/$\psi$ production is measured at mid, semi-forward, and forward rapidities, showing good agreement with LHCb, H1, ZEUS, and fixed-target experiment results within uncertainties~\cite{PRD108}. Comparisons with theoretical models JMRT (power-law description with dominant NLO corrections)~\cite{JMRT} and CCT (color dipole approach) reveal strong sensitivity to gluon distributions at low-$x$~\cite{CCT}, shown in Fig~\ref{Fig:energydep} (b).
\section{Incoherent and dissociative \Jpsi measurements with ALICE}
To search for gluon saturation, $|t|$-dependent measurements are effective observables. The $|t|$-dependent incoherent photoproduction can be used to study the variance (quantum fluctuations). For heavy nuclei, saturation is expected at higher-$x$, where the photon interacts with the sub-nucleonic structure, making $\sigma_{ \gamma \rm Pb}$ sensitive to quantum fluctuations~\cite{PLB772}. The incoherent J/$\psi$ photonuclear cross section was measured as a function of Mandelstam $|t|$ in Pb-Pb collisions at $\sqrt{s_{NN}}$ = 5.02 TeV. Models incorporating quantum fluctuations of the gluon density describe the data better than those without, although no model fully reproduces the normalization from proton to nuclear targets. The slope of the $|t|$-dependence incoherent cross section provides sensitivity to spatial gluon fluctuations, allowing for the first-time probing of gluonic “hot spots” in Pb nuclei~\cite{PRL132}.

In dissociative processes, the photon interacts with the subnucleonic structure, making $\sigma_{\gamma \rm Pb}$ sensitive to the variance rather than the average of the gluon distribution~\cite{CCT}. Dissociative $\gamma p$ production provides access to subnucleonic fluctuations inside the proton~\cite{PLB772}.
HERA data does not cover the full kinematic range accessible at the LHC due to lower energies. The first measurement at collider energies was performed in p-Pb collisions at $\sqrt{s_{NN}}$ = 8.16 TeV. ALICE results are consistent with H1 data for absolute cross sections and agree well with the CCT “hot spots” model, which predicts a maximum cross section at $W_{\gamma \rm p} \sim $ 500 GeV. These measurements serve as a probe for subnucleonic fluctuations inside the proton~\cite{PRD108}.

\section{UPC physics prospects with ALICE experiment in Run 3 and beyond}

ALICE has undergone a significant upgrade, including new detectors, enhanced readouts, a new Central Trigger Processor, and an upgraded DAQ/Offline system~\cite{Aliceupgrade, Alic3upc}.
The Forward Calorimeter (FoCal) detector, planned for Run 4 (starting in 2029), will be positioned 7 m from IP2 on the on the opposite side of the muon arm, covering  $3.4 < \eta < 5.8 $~\cite{focal}.

\begin{figure}[htb]
\includegraphics[width=0.52\textwidth]{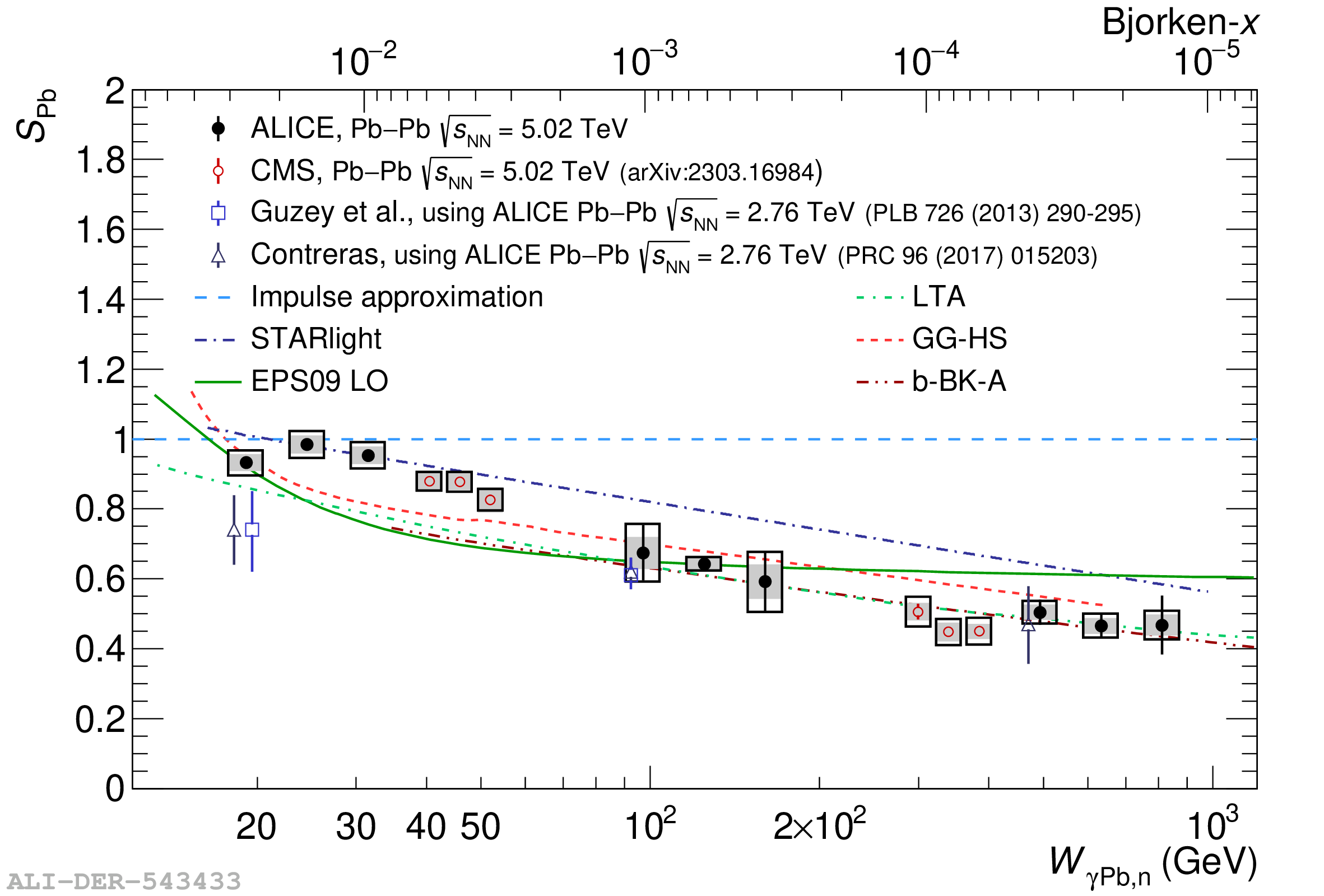}
\includegraphics[width=0.48\textwidth]{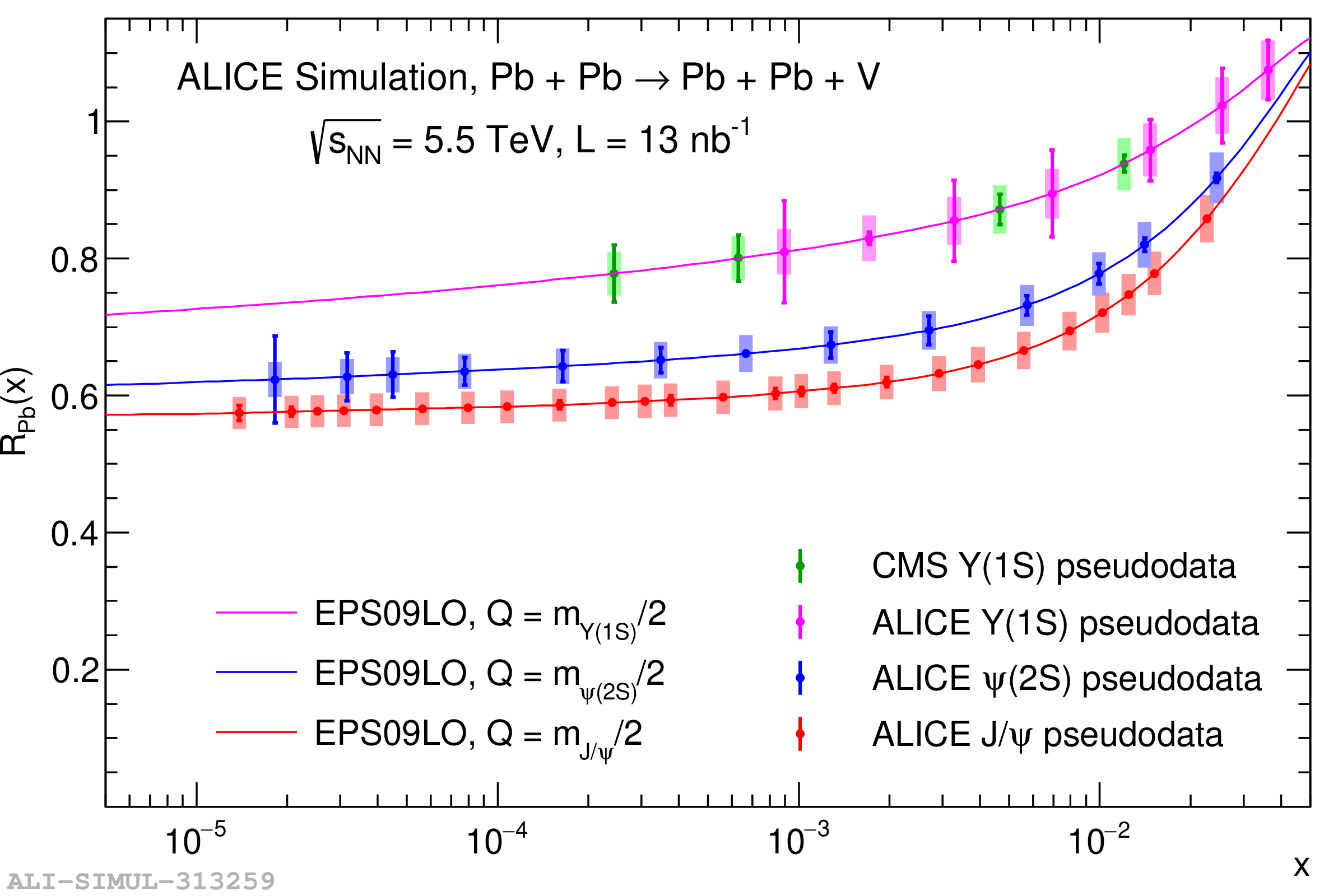}
\caption{(a). Nuclear suppression factor of coherent J/$\psi$ cross-section as a function of energy compared with various theoretical models in Run 2~\cite{JHEP10}. (b). Run 3/4 projection of nuclear suppression factor of coherent J/$\psi$, $\psi(2S)$, $\Upsilon$ cross-sections~\cite{yellowrep}.  }
\label{Fig:supfactor}
\end{figure}







In Run 3 and 4, the increased integrated luminosity enables precision studies of vector meson photoproduction in UPCs, reducing uncertainties for nuclear suppression factors to $\sim 4\%$~\cite{yellowrep}, Fig~\ref{Fig:supfactor} (a). New opportunities include double differential studies, double vector meson photoproduction, UPC bottomonia production, and dissociative J/$\psi$ in Run 3 with FoCal acceptance in Run 4. Run 3 might also allow exclusive J/$\psi$ and $\psi(2S)$ studies in p-Pb UPCs, with further enhancements expected in Run 4 using FoCal acceptance as projected in this article~\cite{JDdissocuative}. 

\end{document}